\documentclass[prd,aps,showpacs,nofootinbib]{revtex4}
\usepackage{amssymb}
\usepackage{graphicx}
\usepackage{amsmath}

\newcommand{\be}{\begin{equation}}
\newcommand{\ee}{\end{equation}}
\newcommand{\bq}{\begin{eqnarray}}
\newcommand{\eq}{\end{eqnarray}}

\begin{document}
\title{Lorentz violation with an invariant minimum speed as foundation of the uncertainty
principle in Minkowski, dS and AdS spaces} 
\author{*Cl\'audio Nassif Cruz}

\affiliation{CPFT-MG:Centro de Pesquisas em F\'isica Te\'orica-MG, Rua Rio de Janeiro 1186, Lourdes, Belo Horizonte-MG, CEP:30.160-041, Brazil.\\
claudionassif@yahoo.com.br} 

\begin{abstract}
This research aims to provide the geometrical foundation of the uncertainty principle within a new causal structure of spacetime so-called Symmetrical Special Relativity (SSR), where there emerges a Lorentz violation due to the presence of an invariant minimum speed 
$V$ related to the vacuum energy. SSR predicts that a dS-scenario occurs only for a certain regime of speeds $v$, where $v<v_0=\sqrt{cV}$, which represents the negative gravitational potentials ($\Phi<0$) connected to the cosmological parameter $\Lambda>0$. For $v=v_0$, Minkowski (pseudo-Euclidian) space is recovered for representing the flat space 
($\Lambda=0$), and for $v>v_0$ ($\Phi>0$), Anti-de Sitter (AdS) scenario prevails 
($\Lambda<0$). The fact that the current universe is flat as its average density of matter distribution ($\rho_m$ given for a slightly negative curvature $R$) coincides with its vacuum energy density ($\rho_{\Lambda}$ given for a slightly positive curvature 
$\Lambda$), i.e., the {\it cosmic coincidence problem}, is now addressed by SSR. SSR provides its energy-momentum tensor of perfect fluid, leading to the EOS of vacuum ($p=-\rho_{\Lambda}$). Einstein equation for vacuum given by such SSR approach allows us to obtain $\rho_{\Lambda}$ associated with a scalar curvature $\Lambda$, whereas the solution of Einstein equation only in the presence of a homogeneous distribution of matter $\rho_m$ for the whole universe presents a scalar curvature $R$, in such a way that the presence of the background field $\Lambda$ opposes the Riemannian curvature $R$, thus leading to a current effective curvature $R_{eff}=R+\Lambda\approx 0$ according to observations. This corrects the notion of gravity as being only of Riemannian origin as the flat space has connection with a background gravity. In view of the current dS-scenario with a quasi-zero $\Lambda$ slightly larger than $\left|R\right|$, we will just obtain a Generalized Uncertainty Principle (GUP) given in the cases of weak gravity and anti-gravity. 
\end{abstract}

\maketitle

\section{Introduction}

Lorentz violations can be observable in nature\cite{Bluhm,Jackiv}. In the last
two decades, the physicists have shown a great interest in the theories that contained the breakdown of Lorentz symmetry in many scenarios\cite{Carroll,Kostelecky,Kostelecky1, Jacobson,Jun,Almeida,Schreck,Tiberio,Belich,Perennes,Lambiase} and also the so-called Deformed Special Relativities (DSR)\cite{Smolin,Camelia}, in spite of no relevant experimental fact has demonstrated the existence of a Lorentz violation until the present time. However, there could be the evidence that the Lorentz symmetry breaking may exist in a very low energy regime due to the presence of a vacuum energy density connected to the well-known cosmological constant $\Lambda$ related to a universal background field (a preferred 
reference frame) associated to an invariant minimum speed. So the effects of an invariant
minimum speed on the breakdown of Lorentz symmetry at lower energies have already been exaustively investigated by the so-called Symmetrical Special Relativity (SSR) theory, where there should be two invariant speeds, namely the well-known speed of light $c$ 
(the maximum speed) for much higher energies and an invariant minimum speed 
$V(\sim 10^{-14}m/s)$ for much lower energies, leading to kinematic non-locality\cite{N2016,N2015,N2008,Rodrigo,N2018,N2010,N2012,acus2,tachyon}. 

Here we should mention an interesting recent paper about SSR entitled ``Lorentz violation with a universal minimum speed as foundation of de Sitter relativity''\cite{Rodrigo}, where
we have shown that the invariant minimum speed provides the foundation for understanding
the conformal metric that represents the dS-relativity. So, we have explored the nature of the SSR-metric\cite{N2016} in order to understand the origin of the conformal factor that appears in the metric by deforming Minkowski metric by means of a scale factor $\Theta(v)$\cite{Rodrigo} depending on the minimum speed that breaks down the Lorentz symmetry, thus leading to a positive cosmological constant. We have also found that SSR-metric provides a set of infinite curvatures\cite{Rodrigo}, i.e., it behaves like an extended dS-relativity, where a strong anti-gravity can be also taken into account when $\Lambda$ is too large. In sum, we have shown that SSR-metric includes the following interval of $\Lambda$, namely 
$0<\Lambda<\infty$, such that we have an extended dS-relativity. Furthermore, we have also shown that SSR-metric is a solution of Einstein equation in a dS-Scenario (with positive cosmological constant) governed by vacuum\cite{Rodrigo}, i.e., in the absence of matter ($T_{\mu\nu}=0$). 

In the present paper based on the previous one\cite{Rodrigo}, thanks to the fact that it 
was already shown that SSR-metric is a solution of Einstein equation in dS-scenario, 
it will be possible to use the energy-momentum tensor $T_{\mu\nu}^{vac}=T_{\mu\nu}^{ssr}$ of perfect fluid (vacuum) given by the $4$-velocities of SSR in the place of the usual 
$T_{\mu\nu}$ that presents the well-known $4$-velocities with Lorentz invariance. So we will go beyond by replacing Einstein equation with the cosmological constant and absence of matter by its equivalent form given in the SSR-vacuum, where the {\it ad-hoc} term of cosmological constant is replaced by its equivalent tensor $T_{\mu\nu}^{ssr}$ of vacuum with the $4$-velocities of SSR\cite{N2016}. After this, we will make the limit $v\rightarrow V$ in the second term of the equation according to SSR. This is exactly equivalent of making $p=-\rho$, i.e., the equation of state (EOS) of vacuum (the cosmological constant). 

Finally, by comparing such term of vacuum of SSR with the well-known {\it ad hoc} term of cosmological constant in dS-scenario, we will find $\rho_{\Lambda}$, which is consistent with the result of the modern cosmology of an accelerated universe, where the cosmic coincidence 
$\rho_{\Lambda}\approx\rho_{matter}$ leads to a flat universe as it will be shown in Section 4, contrary to a Riemannian (curved) universe predicted by the standard cosmology. In this 
sense, we should realize that the geometrical foundations of the standard cosmology are not equivalent geometrical representations of the observations, because Riemannian geometry lacks some details in the description of the universe. In fact, it is notorious that there is a  topological deficiency that cannot be ignored in Einstein's gravitational theory because the flat spacetime acts also as a ground state for gravitation and SSR is able to remove such 
lack, since the cosmological constant $\Lambda$ as an impediment to the existence of Minkowski spacetime is due to its direct connection with the minimum speed 
$V$\cite{N2016,N2015}, so that SSR-spacetime (SSR-conformal metric\cite{Rodrigo}) is a solution of Einstein's equation with the presence of $\Lambda$ and absence of matter, representing a new ground state of the gravitational field, i.e., the de-Sitter (dS) ground state. 

In Section 2, we will make a brief review of the transformations of spacetime in SSR\cite{N2016}.

In Section 3, we will investigate the origin of the Uncertainty Principle and the undulatory nature of matter within the SSR scenario.

In Section 5, we will make a generalization of the Uncertainty Principle (GUP) in dS and 
AdS-spaces, where just a weak gravity and anti-gravity due to a tiny positive cosmological
constant will be taken into account as the current universe is practically flat according 
to the $\Lambda$CDM model, so that we are just interested to investigate GUP close to 
such $\Lambda$CDM scenario. 

\section{Transformations of spacetime in SSR}

\begin{figure}
\includegraphics[scale=1.2]{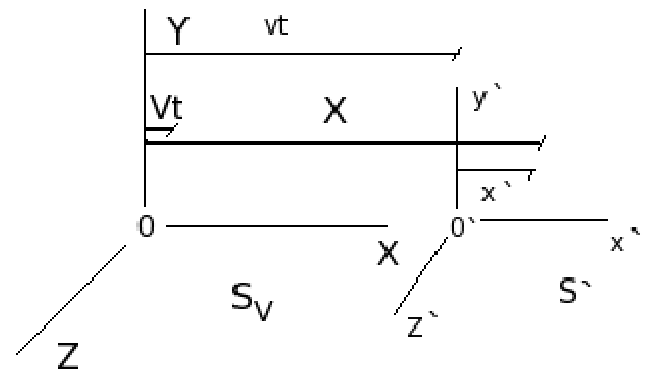}
\caption{In this special case $(1+1)D$, the referential $S^{\prime}$ moves in $x$-direction with a speed $v(>V)$ with respect to a preferred reference frame connected to the ultra-referential $S_V$ (background field). If $V\rightarrow 0$, $S_V$ is eliminated and thus the Galilean frame $S$ takes place, recovering the Lorentz transformations.}
\end{figure}

 The $(1+1)D$-transformations in SSR\cite{N2016,N2015,N2008,Rodrigo,N2018,N2010,N2012} 
 with $\vec v=v_x=v$ (Fig.1) are 

\begin{equation}
x^{\prime}=\Psi(X-vt+Vt)=\theta\gamma(X-vt+Vt) 
\end{equation}

and 

\begin{equation}
t^{\prime}=\Psi\left(t-\frac{vX}{c^2}+\frac{VX}{c^2}\right)=\theta\gamma\left(t-\frac{vX}{c^2}+\frac{VX}{c^2}\right), 
\end{equation}
where the factor $\theta=\sqrt{1-V^2/v^2}$ and $\Psi=\theta\gamma=\sqrt{1-V^2/v^2}/\sqrt{1-v^2/c^2}$. 

The $(3+1)D$-transformations in SSR were shown in a previous paper\cite{N2016} as follows: 

\begin{equation}
\vec r^{\prime}=\theta\left[\vec r + (\gamma-1)\frac{(\vec r.\vec v)}{v^2}\vec v-\gamma\vec v(1-\alpha)t\right], 
\end{equation}
where $\alpha=V/v$. 

And 

\begin{equation}
t^{\prime}=\theta\gamma\left[t-\frac{\vec r.\vec v}{c^2}(1-\alpha)\right]. 
\end{equation}

From Eq.(3) and Eq.(4), we can verify that, if we consider $\vec v$ to be in the same direction of $\vec r$, with $r=X$, we recover 
the special case of $(1+1)D$-transformations given by Eq.(1) and Eq.(2). 

The metric of SSR is a Minkowski metric deformed by a multiplicative function, i.e., a scale factor $\Theta(v)$ with $v$-dependence. $\Theta(v)$ works like a conformal factor, which leads to a kind of dS-metric\cite{Rodrigo}, namely $d\mathcal S^{2}=\Theta\eta_{\mu\nu}dx^{\mu}dx^{\nu}$, where $\Theta=\Theta(v)=\theta^{-2}=1/(1-V^2/v^2)\equiv 1/(1-\Lambda r^2/6c^2)^2$\cite{Rodrigo} is the conformal factor and $\eta_{\mu\nu}$ is the well-known
Minkowski metric. 

\section{Geometrization of the fundamental quantum phenomena within SSR scenario}

This section aims to investigate the geometrical origin of the Uncertainty Principle\cite{N2012} and the undulatory nature of matter within a quantum 
Machian scenario where the vacuum energy of the background framework $S_V$\cite{N2016} plays a fundamental role in understanding the inertial mass
of a particle, which has origin in its interaction with the whole universe by means of the vacuum-$S_V$. In this sense, we should consider such interaction as being related to a quantum Machian scenario generated by the spacetime of SSR.    

\subsection{The concepts of reciprocal $4$-velocity and reciprocal $4$-momentum} 

The idea of reciprocal speed $v_{rec}=v_0^2/v=cV/v$\cite{N2012,acus2,tachyon} allows us to build the contravariant and covariant $4$-reciprocal velocities. As we already know that the contravariant and covariant $4$-velocities are $\mathcal U^{\mu}=\Psi(v)\left[1,v_{\alpha}/c\right]$ and $\mathcal U_{\mu}=\Psi(v)\left[1,-v_{\alpha}/c\right]$\cite{N2016}, we simply write the contravariant and covariant $4$-reciprocal velocities in the following way: 

\begin{equation}
 \mathcal U^{\mu}_{rec}=\Psi(v_{rec})\left[1,\frac{v_{rec,\alpha}}{c}\right]=
 \left[\frac{\sqrt{1-\frac{V^2}{v_{rec}^2}}}{\sqrt{1-\frac{v_{rec}^2}{c^2}}},~
  \frac{v_{rec,\alpha}\sqrt{1-\frac{V^2}{v_{rec}^2}}}{c\sqrt{1-\frac{v_{rec}^2}{c^2}}}\right]  
\end{equation}

and 

\begin{equation}
 \mathcal U_{\mu,rec}=\Psi(v_{rec})\left[1,-\frac{v_{rec,\alpha}}{c}\right]=
 \left[\frac{\sqrt{1-\frac{V^2}{v_{rec}^2}}}{\sqrt{1-\frac{v_{rec}^2}{c^2}}},~
  -\frac{v_{rec,\alpha}\sqrt{1-\frac{V^2}{v_{rec}^2}}}{c\sqrt{1-\frac{v_{rec}^2}{c^2}}}\right],   
\end{equation}
where $\mu=0,1,2,3$ and $\alpha=1,2,3$. We find $v^2_{rec}/c^2=V^2/v^2$ and 
$V^2/v_{rec}^2=v^2/c^2$\cite{N2012}, so that we have $V<v<c$ and $V<v_{rec}<c$, where $V$ is the reciprocal of $c$ and vice versa.

  So, now by substituting $v_{rec}=v_0^2/v=cV/v$ in Eq.(5) and Eq.(6) and after performing the calculations, we finally obtain the contravariant and covariant $4$-reciprocal velocities in function of $v$, namely:

\begin{equation}
\mathcal U^{\mu}_{rec}=\Psi(v)^{-1}\left[1,\frac{V}{v_{\alpha}}\right] 
\end{equation}

and 

\begin{equation}
\mathcal U_{\mu,rec}=\Psi(v)^{-1}\left[1,-\frac{V}{v_{\alpha}}\right], 
\end{equation}

where $\Psi(v)^{-1}=\sqrt{1-v^2/c^2}/\sqrt{1-V^2/v^2}$.

It is important to note that all the components of both contravariant and covariant
$4$-reciprocal velocities diverge when the speed $v$ is closer to the minimum speed $V$,
i.e., $v\rightarrow V$, thus leading to $\mathcal U^{\mu}_{rec}(V)=[\infty,\infty]$ 
and $\mathcal U_{\mu,rec}(V)=[\infty,-\infty]$, while, on the other hand, the components 
of both contravariant and covariant $4$-velocities become zero, i.e., 
$\mathcal U^{\mu}(V)=\mathcal U_{\mu}(V)=[0,0]$. Thus we realize that the reciprocal 
$4$-velocity represents exactly the inverse of the $4$-velocity in the limit 
$v\rightarrow V$. This leads us to strengthen the conclusion that the spacetime of SSR naturally contains a quantum aspect of non-locality given by an ``internal'' motion
($\mathcal U^{\mu}_{rec}$) of the particle that diverges when its momentum tends to zero. 

 We will show that the divergence of the reciprocal $4$-velocity $\mathcal U^{\mu}_{rec}$
so close to the minimum speed $V$ is directly related to a complete delocalization of 
the particle when its momentum goes to zero so close to $V$. This means the increasing of
the uncertainty on position of a particle when the uncertainty on momentum tends to zero,
according to the Uncertainty Principle, but this subject will be deeper investigated soon.

  The contravariant and covariant reciprocal $4$-momenta are given as follows: 
  
  \begin{equation}
 \mathcal P^{\mu}_{rec}=m_0c\mathcal U^{\mu}_{rec}=
 m_0c\Psi(v)^{-1}\left[1,\frac{V}{v_{\alpha}}\right] 
  \end{equation}

 and 
 
 \begin{equation}
 \mathcal P_{\mu,rec}=m_0c\mathcal U_{\mu,rec}=
 m_0c\Psi(v)^{-1}\left[1,-\frac{V}{v_{\alpha}}\right]. 
 \end{equation}
 
 Here it must be stressed that the reciprocal $4$-momentum has an important physical 
 meaning. So, in order to perceive such a new physical meaning, let us make the vacuum 
 approximation $v\approx V$ for the reciprocal $4$-momentum and so we obtain  
 
 \begin{equation}
 \mathcal P^{\mu}_{rec} (v\approx V)=
 \left[\frac{m_0c}{\sqrt{1-\frac{V^2}{v^2}}},\frac{m_0}{\sqrt{1-\frac{V^2}{v^2}}}
 \left(\frac{cV}{v}\right)\right],  
 \end{equation}
 where $cV/v=v_0^2/v=v_{rec}$ is the reciprocal speed\cite{N2012}. 
 
 We should note that the mass $m_{dress}=m_0/\sqrt{1-V^2/v^2}$\cite{N2015} that appears in the components of the reciprocal $4$-momentum for the approximation $v\approx V$ [Eq.(11)] is already known as being the dressed mass that increases to infinite in the limit 
 $v\rightarrow\infty$. This means that the particle becomes strongly coupled to the background field of the whole universe (vacuum energy) within a so-called quantum Machian 
 scenario\cite{tachyon}, thus leading to a drastic increasing of its inertial (effective)
 mass working like a dressed mass, since the particle ``diffuses'' for being delocalized in the whole space in such a limit of $v\rightarrow V$. Thus, the reciprocal $4$-momentum in
 Eq.(11) should be denominated as the dressed $4$-momentum written simply as follows:  
 
 \begin{equation}
 \mathcal P^{\mu}_{rec} (v\approx V)=
 \left[m_{dress}c,~ m_{dress}v_{rec}\right],  
 \end{equation}
 where the component $P_{dress}=m_{dress}v_0^2/v=m_{dress}v_{rec}$ represents the dressed
 (spatial) momentum of the particle close to the minimum speed $V$ (vacuum regime). 
 
 The components of temporal and spatial momenta in Eq.(12) assume very large values close
 to $V$, as the dressed mass $m_{dress}$ becomes too large. However, here it must be stressed
 that both components in Eq.(12) or Eq.(11) become equal to each other when $v$  is much
 closer to the minimum speed $V$, that is to say when we make $v\rightarrow V$, we find
 $v_{rec}=c$, such that both components of $\mathcal P^{\mu}_{rec}$ become equal to 
 $m_{rec}c$. So we can accurately write the following limit: 
 
 \begin{equation}
 \mathcal P^{\mu}_{rec} (v\rightarrow V)=
 \left[m_{dress}c,~ m_{dress}c\right]\rightarrow \left[\infty,~ \infty\right].   
 \end{equation}
 
 Eq.(13) shows two fundamental features that describe the particles of vacuum, namely:
 
 a) As both components of the dressed $4$-momentum are in equal-footing, i.e., 
 $P^0_{dress}=P_{dress}=m_{dress}c$, we conclude that the vacuum is made up of extremely giant bosons that carry very high momenta as their masses are too large. 
 
 b) The vacuum is isotropic and the masses of the moving particles have origin in their
 interactions with such bosons that are extremely giant, in such a way that, only for
 $v>>V$, the couplings of the particles with the background bosons become weaker and so the
 particles recover their finite masses $m_0$, since we have 
 $m_{dress}=m_0/\sqrt{1-V^2/v^2}\approx m_0$ for $V/v\approx 0$ or then $v>>V$.
 
  It is known that the Standard Model (SM) predicts the Higgs bosons that are very heavy
($\approx 125GeV/c^2$) and they provide the masses of the particles. However, SM does 
not include gravity and that is the intriguing question still unsolved. As the inertial masses are equivalent to the gravitational masses, we expect that the most fundamental vacuum should have a gravitational origin in order to provide the masses of the particles within a Machian scenario, which is not taken into account in SM. In this sense, it seems that SM is not still the most fundamental theory to explain the origin of the masses as gravity is directly related to the inertia. Thus a theory beyond SM is needed to give us a more fundamental explanation for the origin of inertia. Eq.(13) is beyond SM as it provides 
extremely heavy bosons whose masses can be much larger than the mass of Higgs boson. Such
extremely heavy bosons are due to the existence of a gravitational vacuum\cite{N2016} associated with a background field that breaks Lorentz symmetry within a Machian scenario
\cite{tachyon}. Here is the novelty that is beyond SM. So we hope LHC be able to produce bosons much heavier than the Higgs bosons at much higher energies. This is a great challenge
that aims to search for the incompleteness of SM due to the background field of SSR by breaking the Lorentz symmetry. 

\subsection{Geometrical description of the Uncertainty Principle within SSR scenario}

Before making the geometrical description of the Uncertainty Principle, let us rewrite 
the four $4$-vectors, as follows:

  \begin{equation}
  \mathcal P^{\mu}=m_0c\Psi(v)\left[1,\frac{v_{\alpha}}{c}\right];  
  \end{equation}

 \begin{equation}
  \mathcal P_{\mu}=m_0c\Psi(v)\left[1,-\frac{v_{\alpha}}{c}\right];  
 \end{equation}

 \begin{equation}
 \mathcal U^{\mu}_{rec}=\Psi(v)^{-1}\left[1,\frac{V}{v_{\alpha}}\right];  
 \end{equation}

 and 

 \begin{equation}
 \mathcal U_{\mu,rec}=\Psi(v)^{-1}\left[1,-\frac{V}{v_{\alpha}}\right],   
 \end{equation}
 where $\mu=0,1.2,3$ and $\alpha=1,2,3$. 
 
 By first making the scalar product of Eq.(15) by Eq.(16), we obtain 
 
 \begin{equation}
 \mathcal P_{\mu}\mathcal U^{\mu}_{rec}=m_0c\left(1-\frac{V}{c}\right). 
 \end{equation}
 
As the ratio $V/c=\xi(\sim 10^{-22})$\cite{N2016} is too small, we can neglect $\xi(<<1)$
in Eq.(18) and simply write 

\begin{equation}
 \mathcal P_{\mu}\mathcal U^{\mu}_{rec}=m_0c. 
\end{equation}

Now it is important to notice that each particle with mass $m_0$ has its own reduced Compton
wavelength $\lambdabar_C$, such that the universal reduced Planck constant $\hbar$ can be 
always written in the following way: 

\begin{equation}
 m_0c\lambdabar_C=\hbar,
\end{equation}
where $m_0$ and $\lambdabar_C$ represent respectively the mass and the reduced Compton wavelength of any particles (e.g: for the electron, we have $m_0=m_e\sim 10^{-30}Kg$ with 
$\lambdabar_C=\lambdabar_{Ce}\sim 10^{-12}m$). 

So, according to Eq.(20), by multiplying both terms of Eq.(19) by $\lambdabar_C$, we find

\begin{equation}
 \mathcal P_{\mu}(\lambdabar_C\mathcal U^{\mu}_{rec})=\hbar. 
\end{equation}

From Eq.(21), we can take the following reciprocal contravariant $4$-vector, namely:

\begin{equation}
 \mathcal X^{\mu}_{rec}=\lambdabar_C\mathcal U^{\mu}_{rec}. 
\end{equation}
 
 In order to understand the important meaning of the $4$-vector Eq.(22), let us first substitute Eq.(16) in Eq.(22) and so we write
 
 \begin{equation}
 \mathcal X^{\mu}_{rec}=\frac{\sqrt{1-\frac{v^2}{c^2}}}{\sqrt{1-\frac{V^2}{v^2}}}\left[\lambdabar_C,\frac{V}{v_{\alpha}}\lambdabar_C\right]. 
 \end{equation}
 
 As the components of $\mathcal X^{\mu}_{rec}$ [Eq.(23)] have dimension of length, then its components provide a deformation of wavelength that goes to the infinite when 
 $v\rightarrow V$, i.e., $\mathcal X^{\mu}_{rec}(V)=[\infty,\infty]$. Thus we conclude that such $4$-vector presents quantum nature by representing exactly a complete delocalization of the particle (uncertainty on position) close to $V$ when the components of its $4$-momentum
 $\mathcal P_{\mu}$ go to zero in such a limit, i.e.,$\mathcal P_{\mu}(V)=[0,0]$. In view of this, the $4$-vector in Eq.(23) is denominated as the {\it delocalization $4$-vector} in SSR spacetime. 
 
On the other hand, Eq.(23) shows that the particle becomes much better located when 
$v$ is too close to $c$, i.e., $\mathcal X^{\mu}_{rec}(c)=[0,0]$, while 
$\mathcal P_{\mu}(c)=[\infty,-\infty]$ in Eq.(15). 
 
As we have shown that Eq.(23) represents the delocalization $4$-vector associated with 
the uncertainty on position, Eq.(21) is simply written as 

\begin{equation}
\mathcal P_{\mu}\mathcal X^{\mu}_{rec}=\hbar, 
\end{equation}
which is the $4$-vector form of the momentum-position uncertainty in SSR spacetime. 

We can alternatively get Eq.(24) within a metric form by taking into account the
metric $\eta^{\mu\nu}=(1,-1,-1,-1)$, so that we write:

\begin{equation}
\eta^{\mu\nu}\mathcal P_{\mu}\mathcal X_{\nu,rec}=\hbar. 
\end{equation}

Secondly by making the scalar product of Eq.(14) by Eq.(17), we obtain essentially the same
previous results, but now having changed the signal in the spatial components of the $4$-vectors since we simply change contravariant vector by the covariant vector and vice-versa. 
Thus we find 

\begin{equation}
 \mathcal P^{\mu}(\lambdabar_C\mathcal U_{\mu,rec})=\hbar,  
\end{equation}
from where we take the reciprocal covariant $4$-vector (delocalization $4$-vector), namely:

\begin{equation}
\mathcal X_{\mu,rec}=\lambdabar_C\mathcal U_{\mu,rec}. 
\end{equation}

So by substituting Eq.(17) in Eq.(27), now we obtain the delocalization $4$-vector in its
covariant form as follows:

\begin{equation}
 \mathcal X_{\mu,rec}=\frac{\sqrt{1-\frac{v^2}{c^2}}}{\sqrt{1-\frac{V^2}{v^2}}}\left[\lambdabar_C, -\frac{V}{v_{\alpha}}\lambdabar_C\right],  
 \end{equation}
so that we simply write Eq.(26) as follows:

\begin{equation}
\mathcal P^{\mu}\mathcal X_{\mu,rec}=\hbar, 
\end{equation}
which is essentially the same result given in Eq.(24), but we just change the covariant
form by the contravariant one and vice-versa. 

Now we can also get Eq.(29) within a metric form by considering the covariant metric 
$\eta_{\mu\nu}=(1,-1,-1,-1)$, so that we write:

\begin{equation}
\eta_{\mu\nu}\mathcal P^{\mu}\mathcal X^{\nu}_{rec}=\hbar. 
\end{equation}

Eq.(30) and Eq.(25) represent the same relation, but given in its covariant and contravariant forms. 

It is interesting to note that the energy-time uncertainty relation contains only scalar
quantities. Such uncertainty relation naturally appears inside Eq.(24) or Eq.(29), written
in the following alternative way: 

\begin{equation}
\mathcal P_{\mu}\mathcal X^{\mu}_{rec}=\mathcal P^{\mu}\mathcal X_{\mu,rec}=
m_0c\lambdabar_C=m_0c^2\Psi\left(\frac{\lambdabar_C}{c}\right)\Psi^{-1}=\hbar, 
\end{equation}
where $E=m_0c^2\Psi$ and $\lambdabar_C/c=t_C$ is the reduced Compton time with which the speed of light takes to travel the reduced Compton length of the particle. 

From Eq.(31), if we make $v\rightarrow V$, we find an infinitely deformed time 
$\tau=t_C\Psi^{-1}(v\approx V)\rightarrow\infty$, such that we obtain
$E=m_0c^2\Psi(v\approx V)\rightarrow 0$. On the other hand, for $v\rightarrow c$, we find
$\tau\rightarrow 0$ and $E\rightarrow\infty$. 

So, from Eq.(31), let us get the following energy-time relation:

\begin{equation}
E\tau=\hbar,
\end{equation}
where $E=E_0\Psi$ and $\tau=t_C\Psi^{-1}$, with $t_C=\lambdabar_C/c$.

It is very important to stress that the speed $v$ in all the equations of SSR is given with 
respect to a preferred reference frame $S_V$ related to the unattainable minimum
speed $V$ (Fig.1). So, all those equations are observer-independent as 
there is no observer at $S_V$. This is the reason why the relations given in Eq.(24), 
Eq.(25), Eq.(29), Eq.(30) and Eq.(32) provide exactly the minimum value of ``uncertainty'', 
i.e., the reduced Planck constant $\hbar$. Actually, the constant $\hbar$ represents 
a minimum quantum action inherent in SSR spacetime, working like an intrinsec 
``uncertainty'', which is observer-independent within an objective reality. However, when a classical observer is taken into account at any Galilean (inertial) reference frame $S$ 
($v=0$), it was already shown that both energy $E$ and momentum $P$ with respect to $S_V$
(no observer) are interpreted as uncertainties $\Delta E$ and $\Delta p$ at any Galilean framework, i.e., we have shown that $E\equiv\Delta E$ and $P\equiv\Delta p$\cite{N2012}. It was also shown that $X\equiv\Delta x$ and $\tau\equiv\Delta t$\cite{N2012}, so that Eq.(24) or Eq.(29) and Eq.(32) given with respect to $S_V$ are interpreted as the minimum uncertainties of momentum-position and energy-time for a classical observer at a Galilean reference frame $S$, written as follows: 

\begin{equation}
[\mathcal P_{\mu}\mathcal X^{\mu}_{rec}]_{S_V}=
[\mathcal P^{\mu}\mathcal X_{\mu,rec}]_{S_V}\equiv[\Delta p\Delta x]_S=\hbar 
\end{equation}

and 

\begin{equation}
[E\tau]_{S_V}\equiv[\Delta E\Delta t]_S=\hbar. 
\end{equation}

We should realize that Eq.(33) and Eq.(34) do not still take into account the presence of an observer that try to measure the quantities $P$, $X$, $E$ and $\tau$ by emitting a photon that transfers momentum and energy to the particle, so that there emerges an increasing of the uncertainties of such quantities, thus leading to the well-known inequalities, namely
$\Delta p\Delta x\geq\hbar$ and $\Delta E\Delta t\geq\hbar$.

\subsection{The concept of reciprocal metric in SSR and its connection with the undulatory
nature of particles}

 When performing the scalar product between the delocalization $4$-vectors given in Eq.(23)
 and Eq.(28), we find
 
 \begin{equation}
 \mathcal X^{\mu}_{rec}\mathcal X_{\mu,rec}=\lambdabar_C^2\left(1-\frac{v^2}{c^2}\right), 
 \end{equation}
 
 or then
 
 \begin{equation}
 \frac{1}{\left(1-\frac{v^2}{c^2}\right)} 
 \eta^{\mu\nu}\mathcal X_{\mu,rec}\mathcal X_{\nu,rec}=\lambdabar_C^2. 
 \end{equation}
 
It is very important to realize that Eq.(36) represents the SSR reciprocal metric. The meaning of the reciprocal metric is the great novelty of SSR as it provides a geometrical description of the undulatory nature of particles by means of their delocalization 
$4$-vectors, and where the squares of their Compton wavelengths are invariant as well as is the square of spacetime interval ($S^2$) in SSR metric, i.e., 

\begin{equation}
 \frac{1}{\left(1-\frac{V^2}{v^2}\right)} 
 \eta^{\mu\nu}\mathcal X_{\mu}\mathcal X_{\nu}=S^2, 
\end{equation}
where $\mathcal X_{\mu}$ is the position $4$-vector. 

As we already know that the metric tensor of SSR in Eq.(37) is  
$G^{\mu\nu}=[1/(1-V^2/v^2)]\eta^{\mu\nu}$, it is interesting to verify that the reciprocal 
metric tensor is written as

\begin{equation}
 G^{\mu\nu}_{rec}=\frac{1}{\left(1-\frac{V^2}{v^2_{rec}}\right)}\eta^{\mu\nu}=
 \frac{1}{\left(1-\frac{v^2}{c^2}\right)}\eta^{\mu\nu},
\end{equation}
where we have $v_{rec}=(cV)/v$.

We should note that the reciprocal metric obtained in Eq.(38) is in fact the own reciprocal metric given in Eq.(36), which is already expected. So let us write Eq.(36) as follows:

\begin{equation}
 \lambdabar_C^2=G^{\mu\nu}_{rec}\mathcal X_{\mu,rec}\mathcal X_{\nu,rec},  
\end{equation}
 which could be denominated as the {\it undulatory metric equation} of a particle. 
 
 Until now we have obtained three fundamental metric equations of SSR, namely 
the first equation is Eq.(25) [or Eq.(30)] that provides the geometrical description of the Uncertainty Principle; the second one is Eq.(36) [or Eq.(39)] that represents the undulatory metric equation of a particle by providing the geometrical description of the wave-particle duality associated with de-Broglie relation and the third one is Eq.(37) that represents 
the conformal metric of SSR.

 Finally the fourth metric equation can be easily obtained from Eq.(24) [or Eq.(29)] by changing the non-reciprocal quantity by its reciprocal form and vice-versa. To do that,
 we must use Eq.(9) [or Eq.(10)], and so we write 
 
 \begin{equation}
\displaystyle\mathcal P^{\mu}_{rec}\mathcal X_{\mu}=
m_0c\Psi(v)^{-1}\Psi(v)(1,V/v) 
\begin{pmatrix}
ct  \\
-X 
\end{pmatrix},
\end{equation}
from where we find

\begin{equation}
\mathcal P^{\mu}_{rec}\mathcal X_{\mu}=m_0c\left(ct-\frac{VX}{v}\right)=
m_0c^2t\left(1-\frac{VX}{vct}\right). 
\end{equation}

But if we admit $t=t_C=\lambdabar_C/c$ in Eq.(41), we would obtain 

\begin{equation}
\mathcal P^{\mu}_{rec}\mathcal X_{\mu}=\hbar\left(1-\alpha\frac{X}{\lambdabar_C}\right), 
\end{equation}
where we have $\hbar=m_0c\lambdabar_C$ and $\alpha=V/v$. 

Eq.(42) can be written in the metric form, namely:

\begin{equation}
\eta_{\mu\nu}\mathcal P^{\mu}_{rec}\mathcal X^{\nu}=
\hbar\left(1-\alpha\frac{X}{\lambdabar_C}\right).  
\end{equation}

Eq.(43) is another novelty and it leads to a deformed uncertainty relation for the reciprocal momentum as the reduced Planck constant is corrected by $v$ and $X$. Although this result deserves to be deeply investigated elsewhere, here let us search for some interpretations of Eq.(43) by first taking into account lower energies when the speed is close to $V$ (vacuum), i.e., $v\approx V$. In this regime, $\mathcal P^{\mu}_{rec}$ is the dressed $4$-momentum
[Eq.(11)] that diverges when $v\rightarrow V$ as the particle becomes strongly coupled to 
vacuum at much lower energies. So, by considering $\alpha\approx 1$ ($v\approx V$) in 
Eq.(43), we find

\begin{equation}
\eta_{\mu\nu}\mathcal P^{\mu}_{rec}(v\approx V)\mathcal X^{\nu}=
\hbar\left(1-\frac{X}{\lambdabar_C}\right),
\end{equation}
where we must have $X\leq ct_C(=\lambdabar_C)$ in such a regime. If $X=ct_C$, we 
would have a light-like interval as $X_{\mu}X^{\mu}=X^2-c^2t^2_C=0$. In this very special 
case of so low energies and with all the points connected by light inside a sphere of 
radius $\lambdabar_C$, the uncertainty relation for the dressed momentum is completely 
violated, so that we find

\begin{equation}
\eta_{\mu\nu}\mathcal P^{\mu}_{rec}(v\approx V)\mathcal X^{\nu}=0. 
\end{equation}

Eq.(45) shows us a particle with so low energy and a too high dressed momentum strongly confined by vacuum inside a very small sphere of radius $\lambdabar_C$. This special phenomenon of confinement could be understood as a quark confinement at too low energies inside the proton (QCD at infrared regime), where the proton radius ($r_p\sim 10^{-15}$m) is exactly in the order of magnitude of the reduced Compton wavelength ($\lambdabar_{Cq}$) for the constituent mass of each quark, namely $m_q=m_{quark}=(1/3)m_p\sim 10^{-27}$ Kg, so that we indeed verify that $\lambdabar_{Cq}=\hbar/m_qc\sim r_p\sim 10^{-15}$m. 

Thus Eq.(45) shows us that, when the quark becomes even much more confined inside the proton just for $v\approx V$, the uncertainty relation is violated in such a way that we could localize the quark inside the proton.

A deeper investigation of the implications of Eq.(43) could show a similarity between QCD-vacuum and the gravitational vacuum within a Machian scenario where the whole universe works like a kind of a big bag inside which each particle presents a certain degree of confinement
given by the reciprocal $4$-momentum. Such similarity between a microscopic confinement system (e.g:quarks in the proton) and the cosmological one (particles in the universe) leads to the so-called holography. 

Other cases emerge from Eq.(43), as for instance, the case of $\alpha<<1$ ($v>>V$) or even 
for $v\rightarrow c$ when the particle becomes unconfined, since 
$\mathcal P^{\mu}_{rec}(v\approx c)\approx 0$. So, due to many interpretations emerging from Eq.(43), such broad issue will be explored elsewhere.

\section {The flat universe and the problem of cosmic coincidence under SSR}

\subsection{The cosmological constant as a positive curvature (cosmological anti-gravity) and its connection with the minimum speed as foundation of the EOS of vacuum}

It is already known that the current  universe has a very low average density of  matter 
$\rho_{m}$ that mysteriously coincides with its vacuum energy density $\rho_{\Lambda}$,  i.e., $\rho_{m}\approx \rho _{\Lambda}\sim 10^{-29}g/cm^3$. This is the well-known {\it cosmic coincidence problem}, which is due to the presence of the tiny value of the cosmological constant  $\Lambda\sim 10^{-35}s^{-2}$ that leads to a cosmological anti-gravity whose repulsive effect counterposes exactly the attractive gravity of the very low density of matter $\rho_{m}$, in such a way that the geometry of the universe becomes quasi-flat according to observations. In this sense the presence of the cosmological constant corrects the topological  deficiency of Einstein equation as the flat space now can be considered a ground state of gravitational field, contrary to Minkowski (flat) space as being a ground state without gravity. 

In this section, we show how the cosmological constant implements such new gravitational ground state in the Einstein equation by means of SSR theory. To do that, we should take into account the energy-momentum tensor of perfect fluid of SSR  ($T_{\mu\nu}^{ssr}$) already investigated in a previous work\cite{N2016}, where we have defined

\begin{equation}
T_{\mu\nu}^{ssr}=(p+\rho)\mathcal U_{\mu}\mathcal U_{\nu}-pg_{\mu\nu},
\end{equation}
where $p$ is the pressure, $\rho$ being the energy density and 
$\mathcal U_{\mu}=\Psi(v)[1,-v_{\alpha}/c]$ ($\alpha=1,2,3$) is the covariant $4$-velocity of the SSR theory, where we already know that $\Psi(v)=\sqrt{1-V^2/v^2}/\sqrt{1-v^2/c^2}$. 

We should have in mind that such energy-momentum tensor given in the limit  of vacuum of SSR, i.e., $v\rightarrow V$, leads to $ T_{\mu\nu,vac}^{ssr}=-pg_{\mu\nu}$, where the first term of $4$-velocities vanishes  as we find $\mathcal U_{\mu}(v\rightarrow V)=0$. Besides this, we note that when considering $p=-\rho$, this leads to the same result, namely:

\begin{equation}
T_{\mu\nu,vac}^{ssr}=-pg_{\mu\nu}=\rho g_{\mu\nu},
\end{equation}
where we realize that the EOS of vacuum ($p=-\rho=-\rho_{\Lambda}$) emerges naturally from the spacetime of SSR in the sense that  $T_{\mu\nu,vac}^{ssr}$ given in the limit $V$ plays the same role of the cosmological  constant  $\Lambda$ placed into the Einstein equation in {\it ad hoc} way for representing  the de-Sitter (dS) scenario. Thus, according to SSR, we write the Einstein equation in a dS scenario (also in the absence of matter), namely:

\begin{equation}
R_{\mu\nu}-\frac{1}{2}Rg_{\mu\nu}=\frac{8\pi G}{c^2}[lim_{v\rightarrow V} T_{\mu\nu}^{ssr}], 
\end{equation}
which leads to 

\begin{equation}
R_{\mu\nu}-\frac{1}{2}Rg_{\mu\nu}=\frac{8\pi G}{c^2}T_{\mu\nu,vac}^{ssr}=
\frac{8\pi G}{c^2}\rho_{\Lambda}g_{\mu\nu},
\end{equation}
or simply 

\begin{equation}
R_{\mu\nu}-\frac{1}{2}Rg_{\mu\nu}-\frac{8\pi G}{c^2}\rho_{\Lambda}g_{\mu\nu}=0,
\end{equation}
with $\rho_{\Lambda}=-p$. 

On the other hand, we already know the Einstein equation in the absence of matter and with cosmological constant in a dS scenario, as follows:  

\begin{equation}
R_{\mu\nu}-\frac{1}{2}Rg_{\mu\nu}-\Lambda g_{\mu\nu}=0.
\end{equation}

Finally, by comparing Eq.(51) above with its equivalent form [Eq.(50)] given in the vacuum of SSR, whose EOS is the same of the cosmological constant ($p=-\rho_{\Lambda}$), we obtain 

\begin{equation}
\rho_{\Lambda}=-p=\frac{\Lambda c^2}{8\pi G},
\end{equation}
or then we write

\begin{equation}
\Lambda=\frac{8\pi G}{c^2}\rho_{\Lambda}, 
\end{equation}
where the pressure is $p<0$ (anti-gravity) and $\Lambda>0$, which represents a dS-space  having positive curvature given by $\Lambda$ (vacuum energy) as we must have 
$\rho_{\Lambda}>0$. 

Although Eq.(53) is already known, the novelty here is that it was obtained by taken into account the vacuum of SSR associated with an invariant minimum speed $V$ that leads naturally to the EOS of vacuum ($p=-\rho_{\Lambda}$) and thus the cosmological constant $\Lambda$, i.e., $\Lambda$ and $\rho_{\Lambda}$ have origin in the spacetime of SSR ($V$). 

\subsection{The negative scalar curvature (attractive gravity) and its connection with the average distribution of matter in the universe}

In this section, we show how the scalar curvature of the universe ($R$) is connected to its  density of matter, generating an attractive gravity that try to oppose its accelerated expansion due to $\Lambda$ in the current universe.  In order to do that, we use the well-known Einstein equation without the cosmological constant and only with the presence of a source of matter (gravity) given by  the  energy-momentum tensor $T_{\mu\nu}$, namely: 

\begin{equation}
R_{\mu\nu}-\frac{1}{2}Rg_{\mu\nu}=\frac{8\pi G}{c^2}T_{\mu\nu}.
\end{equation}

As we are in the cosmological scenario, where we consider the whole universe with its low homogeneous (average) density of matter, the whole universe can be treated as a perfect fluid, including all kinds of matter and also the dark and vacuum energies. In this sense, the  universe is isotropic as it is a perfect fluid with the same pressure $p$ in all directions, so that the tensor $T_{\mu\nu}$ is given by a diagonal matrix. However, in this case, only the effects of gravity (matter) should be taken into account in the energy-momentum tensor of perfect fluid, where we should make a certain approximation such that the effects of matter prevail. 

We should realize that the energy-momentum tensor of perfect fluid of SSR 
($T_{\mu\nu}^{ssr}$) given in Eq.(46) contains matter (gravity regime) and also vacuum energy due to the cosmological constant (anti-gravity regime) whose EOS is obtained just in the vacuum approximation $v\approx V$. So now it is easy to conclude that when making the approximation $v>>V$ in the tensor of SSR [Eq.(46)], the vacuum effects are neglected and the well-known energy-momentum tensor of perfect fluid for representing just matter diluted uniformly in the whole space takes place, so that we write the following equation: 

\begin{equation}
R_{\mu\nu}-\frac{1}{2}Rg_{\mu\nu}=\frac{8\pi G}{c^2}T_{\mu\nu, v>>V}^{ssr}, 
\end{equation}
where $T_{\mu\nu, v>>V}^{ssr}\approx T_{\mu\nu}$, i.e., the energy-momentum tensor of perfect fluid  of General Relativity (GR) theory is recovered. 

According to Eq.(46), we can calculate the trace $T$ of $T_{\mu\nu}^{ssr}$, namely:

\begin{equation}
T=T^{\mu}_{\mu}=(p+\rho)\mathcal U^{\mu}\mathcal U_{\mu}-p\delta^{\mu}_{\mu},
\end{equation}
from where we obtain 

\begin{equation}
T=(p+\rho)\left(1-\frac{V^2}{v^2}\right)-4p,
\end{equation}
where we find $\mathcal U^{\mu}\mathcal U_{\mu}=\mathcal U_{\mu}\mathcal U^{\mu}=(1-V^2/v^2)$ and $\delta^{\mu}_{\mu}=4$. 

It is interesting to note that, when making $v>>V$ in Eq.(57), we obtain the matter and radiation regime far from the vacuum regime, so that we recover the trace $T$ of the well-known energy-momentum tensor of perfect fluid of GR ($T_{\mu\nu}$), namely:

\begin{equation}
T=\rho-3p, 
\end{equation}
where $T=0$ for the case of the electromagnetic field. Here we are only interested in the case of pure matter with $p=0$, so that $T=\rho=\rho_m$. 

 Let us now obtain Eq.(54) [or Eq.(55)] in the form of its mixed components, as follows: 
 
\begin{equation}
R_{\mu}^{\nu}-\frac{1}{2}\delta_{\mu}^{\nu}R=\frac{8\pi G}{c^2}T_{\mu}^{\nu}.
\end{equation}

By making $\mu=\nu$ in Eq.(59) above and knowing that $\delta_{\mu}^{\mu}=4$, we get

\begin{equation}
R=-\frac{8\pi G}{c^2} T,
\end{equation}
where $R$ is the curvature of the whole universe that presents an average density of matter given by  $T=\rho_{matter}=\rho_m$.  So, finally we obtain 

\begin{equation}
R=-\frac{8\pi G}{c^2}\rho_m. 
\end{equation} 

As the trace of any energy-momentum tensor is always positive or null, in the present case  where we just consider the universe filled by matter (including all kinds of attractive matter), we must have $T=\rho_m>0$, so  that we find a negative scalar curvature ($R<0$), which indicates that the presence of matter (attractive gravity) generates a constant negative curvature for representing a maximal space with negative curvature\cite{Rodrigo}, contrary to $\Lambda>0$ [Eq.(53)] that works like a positive curvature (cosmological anti-gravity) that contraposes $R<0$. 

\subsection{The flat space: the $\Lambda$CDM model} 

 It is known that the $\Lambda$CDM model assumes that the universe is composed of photons, neutrinos, ordinary matter (baryons, electrons) and cold (non-relativistic) dark matter (CDM), which only interacts gravitationally, plus dark energy (or a vacuum energy), which is responsible for the observed acceleration in the Hubble expansion. Dark energy is assumed to take the form of a constant vaccuum energy density, referred to as the cosmological constant ($\Lambda$) as shown in Eq.(50) by SSR. The Standard $\Lambda$CDM model with such $6$ parameters further imposes the constraint that space is flat (Euclidean). However we should stress that its exact shape is still a matter of debate in physical cosmology, but experimental data from various independent sources (WMAP, BOOMERanG and Planck) confirm that the observable universe is flat according to $\Lambda$CDM or at least almost flat. It was found about $0.4$\% margin of error whose origin is not clear yet. High-resolution maps of the cosmic microwave background (CMB) radiation has confirmed a flat universe\cite{CMB}, but the presence of such minor errors is still inevitable. 
 
 The universe has about $4$\% of visible matter (vm) given by the baryonic matter, also including electrons since the quantity of anti-matter is minimal and can be neglected. This $4$\% is composed of only about $0.4$\% of stars plus $3.6$\% of interestelar matter (gases), i.e., we write $\Omega_{vm}\approx  0.04$. It has about 
 $23$\% of cold dark matter (CDM), i.e., $\Omega_{CDM}\approx 0.23$ and finally with about $73$\% of dark energy, i.e., $\Omega_{\Lambda}\approx 0.73$, so that the whole universe has $\Omega=\Omega_m+\Omega_{\Lambda}=1$ (a flat universe), where $\Omega_m=\Omega_{vm}+\Omega_{CDM}$ represents the result of all kinds of attractive matter (PS: here the effects of neutrinos and photons ($T=0$) are so minimal that they do not affect our results). Therefore, we can write Eq.(55) by considering all the components of attractive matter, namely: 
 
\begin{equation}
R=-\frac{8\pi G}{c^2}\rho_m=-\frac{8\pi G}{c^2}(\rho_{vm}+\rho_{CDM}). 
\end{equation} 

According to SSR given in the regime of vacuum  ($v\approx V$) whose energy governs about $73$\% of the universe, i.e., $\Omega_{\Lambda}\approx 0.73$ , we have already found

\begin{equation}
\Lambda=\frac{8\pi G}{c^2}\rho_{\Lambda}.
\end{equation}

Finally, by adding the effects of scalar curvatures of repulsive ($\Lambda$) and attractive (vm plus CDM) gravity given in the Eq.(63) and Eq.(62), i.e., $\Lambda+R$, we find the effective scalar curvature $R_{eff}=\Lambda+R$, namely:

\begin{equation}
R_{eff}=R_{vm+\Lambda+CDM}=\frac{8\pi G}{c^2}[\rho_{\Lambda}-(\rho_{vm}+\rho_{CDM})]. 
\end{equation}

As the universe is flat, i.e., $\Omega=1$, then $\Lambda$CDM model becomes consistent with Eq.(64) only if the effective scalar curvature $R_{eff}=R_{vm+\Lambda+CDM}=0$, so that we find 

\begin{equation}
\rho_{\Lambda}-(\rho_{vm}+\rho_{CDM})=0,
\end{equation}
or simply $\rho_{\Lambda}=\rho_{m}$, which is the cosmic coincidence for the current universe, leading to an Euclidian universe as $R_{eff}=0$.

In a previous work\cite{N2016}, according to SSR theory, it was demonstrated that the constant $\Lambda$ is in fact a cosmological parameter that depends on the Hubble radius 
$R_H$, being written as follows:

\begin{equation}
\Lambda=\frac{6c^2}{R_H^2}.
\end{equation}

As $R_H\sim 10^{26}$m, we find $\Lambda\sim 10^{-35}s^{-2}$. This positive tiny order of magnitude of $\Lambda$ is in agreement with observations.  

By substituting Eq.(66) in Eq.(63), we obtain the vacuum energy density $\rho_{\Lambda}$ depending on the Hubble radius, namely: 

\begin{equation}
\rho_{\Lambda}=\frac{3c^4}{4\pi G R_H^2}. 
\end{equation}

On the other hand, we obtain the total average density of matter $\rho_m$ for the spherical universe with total mass $M$ and Hubble radius $R_H$ as being 

\begin{equation}
\rho_m=\frac{3M}{4\pi R_H^3}.
\end{equation}

Due to the cosmic coincidence that leads to a flat universe as $\rho_m=\rho_{\Lambda}$, thus we write

\begin{equation}
\frac{3M}{4\pi R_H^3}=\frac{3c^4}{4\pi G R_H^2}\sim 10^{-29}g/cm^3,
\end{equation}
which indeed occurs for the current Hubble radius $R_H\sim 10^{26}$m, 
where $M=M_{vm}+M_{CDM}\sim 10^{54}kg$. 

Here we should note that, as $\rho_{\Lambda}\propto R_H^{-2}$ decreases slower than $\rho_m\propto R_H^{-3}$ with the Hubble radius expansion, the cosmological anti-gravity of expansion will overcome the attractive gravity among the galaxies,i.e., $\rho_{\Lambda}>\rho_m$ in the future. Therefore SSR predicts that the universe will no longer be flat in the future and it will have an increasing of positive curvature governed by  accelerated expansion, since $R_{eff}=\Lambda+R>0$. So, in the future, SSR theory shows that the 
universe will be governed by a dS-scenario. This is the reason why we are interested in making a generalization of the uncertainty principle (GUP) in a dS-scenario ($\phi<0$)\cite{Rodrigo}, given for the approximation of a weak anti-gravity, i.e., 
$R_{eff}=\Lambda-R=\Lambda_{eff}\geq 0$. In the next section, we will also investigate 
GUP in a AdS-scenario with a negative $\Lambda_{eff}$, but close to zero.  

\section{Generalization of the Uncertainty Principle (GUP) in dS and AdS spaces within 
SSR scenario}

The graph in Fig.2 shows the transition between gravity (AdS-spaces) and anti-gravity
(dS-spaces). 

\begin{figure}     
\centering
\includegraphics[scale=0.46]{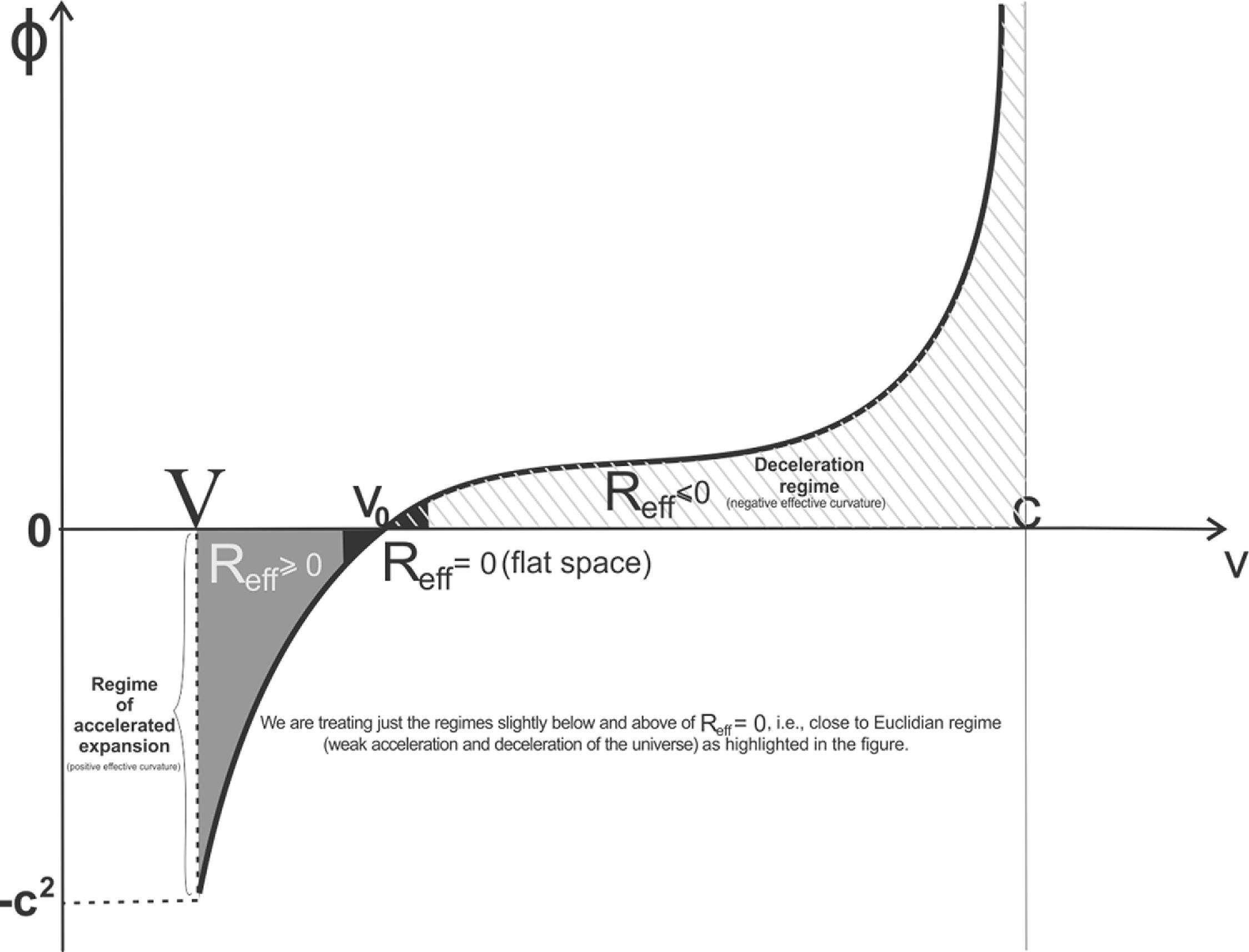}
\caption{The graph shows the scalar potential $\phi(v)=c^{2}\left(\sqrt{\frac{1-\frac{V^2}{v^2}}{1-\frac{v^2}{c^2}}}-1\right)$. It shows the two phases and their connections with 
the effective scalar curvature of the universe ($R_{eff}$), namely the effective gravity 
(right side) with $R_{eff}<0$ and the effective anti-gravity (left site) with 
$R_{eff}>0$. We see the barrier at the right side, representing the relativistic limit close to the speed of light $c$ with $\phi\rightarrow\infty$, and the barrier at the 
left side representing the quantum (anti-gravitational) limit only described by SSR (minimum speed $V$ with $\phi=-c^2$). The intermediary region $V<<v<<c$ is the Newtonian regime, 
which represents a quasi-flat space with a quasi-zero scalar effective curvature, i.e., 
$R_{eff}\approx 0$ as matter and vacuum energy are in equal footing, represeting the state 
of the current universe. The speed $v=v_0=\sqrt{cV}$ is the point of phase transition where
the space is flat ($R_{eff}=0$ with $\phi=0$) below which ($v<v_0$) the repulsive vacuum effect takes place as $R_{eff}>0$ ($\phi<0$), and above which ($v>v_0)$ the attractive 
gravity (matter) takes place as $R_{eff}<0$ ($\phi>0$).} 
\label{Rotulo}
\end{figure}

\subsection{GUP in dS-spaces governed by anti-gravity: $-c^2\leq\phi\leq 0^{-}$ 
 ($R_{eff}\geq 0$)}

From Eq.(18), we have obtained the uncertainty relation in a quasi-flat spacetime by
considering the slight deviation on the reduced Planck constant due to the background 
field at $S_V$, i.e., $\hbar(1-\xi)$ with $\xi=V/c$, which leads exactly to 

\begin{equation}
\eta_{\mu\nu}\mathcal P^{\mu}\mathcal X^{\nu}_{rec}=\hbar(1-\xi). 
\end{equation}

By multiplying both sides of Eq.(70) by the conformal factor $\Theta$, we obtain

\begin{equation}
G_{\mu\nu}\mathcal P^{\mu}\mathcal X^{\nu}_{rec}=
\frac{\hbar(1-\xi)}{\left(1-\frac{V^2}{v^2}\right)}. 
\end{equation}

By expanding the second side of Eq.(71), we write 

\begin{equation}
 G_{\mu\nu}\mathcal P^{\mu}\mathcal X^{\nu}_{rec}=
 \hbar(1-\xi)\left[1+\frac{V^2}{v^2}+\frac{V^4}{2!v^4}+...\right].  
\end{equation}

Let us now consider the approximation for a slightly repulsive potential 
$\phi\approx 0^{-}$, so that $v$ is slightly smaller than $v_0(>>V)$, where $\phi(V)=-c^2$ 
is the most repulsive potential (Fig.2). In doing this in Eq.(72), we write

\begin{equation}
 G_{\mu\nu}\mathcal P^{\mu}\mathcal X^{\nu}_{rec}\approx\hbar(1-\xi)\left[1+\frac{V^2}{v^2}\right].  
\end{equation}

Eq.(73) can be written in its equivalent form of weakly repulsive potential, namely: 

\begin{equation}
 G_{\mu\nu}\mathcal P^{\mu}\mathcal X^{\nu}_{rec}\approx\hbar(1-\xi)
 \left(1-\frac{2\phi}{c^2}\right),  
\end{equation}
where $\phi\leq 0^{-}$ (see Fig.2). 

We already know that the momentum $P$ is given with respect to $S_V$, but it represents 
an uncertainty $\Delta p$ at any Galilean reference frame $S$, i.e., we write 
$P\equiv(\Delta p)_S=\Delta p$\cite{N2012}.  

As we are interested to investigate dS-spaces, where $V<v\leq v_0$ (Fig.2), let us write
the uncertainty $\Delta p$ only for the repulsive sector of gravity (anti-gravity), namely:

\begin{equation}
 \Delta p\equiv P=m_0v\sqrt{1-\frac{V^2}{v^2}}=m_0v\left(1+\frac{\phi(r)}{c^2}\right), 
\end{equation}
where $V<v\leq v_0$ for $-c^2<\phi\leq 0^{-}$. 

 By considering the slightly repulsive potential $\phi\leq 0^{-}$, where $v\leq v_0$
 or then $v\approx v_0^{-}$, so we make the approximation in Eq.(75) as follows: 
 
 \begin{equation}
 \Delta p\approx m_0v_0\left(1+\frac{\phi(r)}{c^2}\right)=
 p_0\left(1+\frac{\phi(r)}{c^2}\right), 
 \end{equation}
 where we simply have made $m_0v\approx m_0v_0=m_0\sqrt{cV}=p_0$, since $v\approx v_0^{-}$. 
 
 From Eq.(76), we get  
 
 \begin{equation}
 \Delta p-p_0\approx\frac{p_0\phi(r)}{c^2},  
 \end{equation}
 from where we obtain 
 
 \begin{equation}
 \frac{2\phi(r)}{c^2}\approx\frac{2(\Delta p-p_0)}{p_0}.   
 \end{equation}
 
 By substituting Eq.(78) in Eq.(74), we finally get 
 
 \begin{equation}
 G_{\mu\nu}\mathcal P^{\mu}\mathcal X^{\nu}_{rec}
 \approx\hbar(1-\xi)\left[1-\frac{2(\Delta p-p_0)}{p_0}\right],  
\end{equation}
where we realize that $\Delta p<p_0$ for dS-spaces with small positive cosmological 
constants (weak anti-gravity).  

 Eq.(79) represents the modified uncertainty relation in the presence of a weakly repulsive
 anti-gravity on the surface of a sphere of dark energy with radius $r$.
 
 \subsection{GUP in AdS-spaces governed by gravity: $0^{+}\leq\phi<\infty$ 
  ($R_{eff}\leq 0$)} 
 
 First of all, it is important to perceive that the well-known conformal 
 metric $(G_{\mu\nu})_{dS}$ [Eq.(71)] given in dS-spaces can be transformed into another
 conformal metric $(G_{\mu\nu})_{AdS}$ given in AdS-spaces only by performing the reciprocal
 form of $(G_{\mu\nu})_{dS}=G_{\mu\nu}$, i.e., we have to calculate the reciprocal metric 
 $G_{\mu\nu, rec}$ in order to obtain $(G_{\mu\nu})_{AdS}$. So we write 
 
 \begin{equation}
  (G_{\mu\nu})_{AdS}=G_{\mu\nu, rec}=
  \frac{1}{\left(1-\frac{V^2}{v_{rec}^2}\right)}\eta_{\mu\nu}. 
 \end{equation}
 
 As we already know that $v_{rec}=v_0^2/v=cV/v$, Eq.(80) is written as 
 
 \begin{equation}
  (G_{\mu\nu})_{AdS}=G_{\mu\nu, rec}=
  \frac{1}{\left(1-\frac{v^2}{c^2}\right)}\eta_{\mu\nu}. 
 \end{equation}
 
 Thus Eq.(71) is transformed into its reciprocal form (AdS-spaces), namely: 
 
 \begin{equation}
 G_{\mu\nu, rec}\mathcal P^{\mu}\mathcal X^{\nu}_{rec}=
 \hbar(1-\xi)\left[1+\frac{v^2}{c^2}+\frac{v^4}{2!c^4}+...\right],  
\end{equation}
which represents the attractive potential ($\phi>0$) as we are in the Lorentz sector, i.e., 
$v_0<v<c$ (Fig.2). 

Let us now consider the approximation for a weakly attractive potential
$\phi\approx 0^{+}$, so that $v$ is not much larger than $v_0$ (Fig.2). In doing this 
in Eq.(82), we write

\begin{equation}
 G_{\mu\nu, rec}\mathcal P^{\mu}\mathcal X^{\nu}_{rec}=
 \hbar(1-\xi)\left[1+\frac{v^2}{c^2}\right].  
\end{equation}

Eq.(83) can now be written in its equivalent form of weakly attractive potential as follows:

\begin{equation}
G_{\mu\nu, rec}\mathcal P^{\mu}\mathcal X^{\nu}_{rec}=
\hbar(1-\xi)\left[1+\frac{2\phi}{c^2}\right],
\end{equation}
where we must have $\phi\geq 0^{+}$. 

As we are now interested to investigate AdS-spaces, where $v_0\leq v<c$ (Fig.2), let us 
write the uncertainty $\Delta p$ only for the attractive sector of gravity (Lorentz sector), namely:

\begin{equation}
 \Delta p\equiv P=\frac{m_0v}{\sqrt{1-\frac{v^2}{c^2}}}=
 m_0v\left(1+\frac{\phi(r)}{c^2}\right), 
\end{equation}
where $0<\phi<\infty$. 

By considering a weakly attractive potential $\phi\geq 0^{+}$, where $v\geq v_0$ or then 
$v\approx v_0^{+}$, so we make the approximation in Eq.(85) as follows:

\begin{equation}
\Delta p\approx p_0\left(1+\frac{\phi(r)}{c^2}\right), 
\end{equation}
where $p_0=m_0v_0$. 

 From Eq.(86), we get  
 
 \begin{equation}
 \Delta p-p_0\approx\frac{p_0\phi(r)}{c^2},  
 \end{equation}
 from where we obtain 
 
 \begin{equation}
 \frac{2\phi(r)}{c^2}\approx\frac{2(\Delta p-p_0)}{p_0}.   
 \end{equation}
 
 By substituting Eq.(88) in Eq.(84), we finally obtain 
 
 \begin{equation}
 G_{\mu\nu, rec}\mathcal P^{\mu}\mathcal X^{\nu}_{rec}
 \approx\hbar(1-\xi)\left[1+\frac{2(\Delta p-p_0)}{p_0}\right],  
\end{equation}
where we have $\Delta p>p_0$ and $G_{\mu\nu, rec}=(G_{\mu\nu})_{AdS}$ for AdS-spaces with
weak gravity. 

Eq.(89) represents the modified uncertainty relation in the presence of a weakly attractive
gravity on the surface of a sphere dominated by matter and with radius $r$.

Here it is interesting to mention a recent reference\cite{GUP} that deals with GUP by
modifying the Heisenberg principle in presence of a gravitational field. The 
authors have found a deformed uncertainty relation which reminds Eq.(89),
since $\Delta p$ also appears in the second side of their relation by modifying 
$\hbar$, thus leading to an effective reduced Planck constant\cite{GUP}.\\\\

{\noindent\bf  Acknowledgements}

I am grateful to the heated discussions with my colleagues Fernando Ant\^onio da Silva, 
Rodrigo Francisco dos Santos and Ant\^onio Carlos Amaro de Faria Jr.

\end{document}